\documentclass[12pt]{iopart}

\usepackage{iopams}  
\usepackage{graphicx}
\usepackage[latin1]{inputenc} 
\usepackage[english]{babel} 
\usepackage[T1]{fontenc}
\usepackage{amssymb}
\usepackage{textcomp}
\usepackage{bm}
\usepackage{color}
\usepackage{tocvsec2}
\newcommand{\PSLstar}{\raisebox{0.5mm}{\small{$\bigstar$}}}

\newcommand{\ket}[1]{\left| #1 \right>} 

\newcommand{\ex}{\mathbf{e_x}}
\newcommand{\ey}{\mathbf{e_y}}

\newcommand{\ez}{\mathbf{e_z}}

\newcommand{\dd}{\mathrm{d}}

\newcommand{\Er}{{E_{\mathrm{r}}}}

\begin{document}

\title[]{Optical cooling and trapping of highly magnetic atoms:
The benefits of a spontaneous spin polarization
}

\author{Davide Dreon, Leonid A. Sidorenkov, Chayma Bouazza, Wilfried Maineult, Jean Dalibard, Sylvain Nascimbene}

\address{Laboratoire Kastler Brossel, Coll\`ege de France, CNRS, ENS-PSL Research University,
UPMC-Sorbonne Universit\'es, 11 place Marcelin Berthelot, 75005 Paris}
\ead{sylvain.nascimbene@lkb.ens.fr}
\begin{abstract}

From the study of long-range-interacting systems to the simulation of gauge fields, open-shell Lanthanide atoms  with their large magnetic moment and narrow optical transitions open novel directions in the field of ultracold quantum gases.  As for other atomic species, the magneto-optical trap (MOT) is the working horse of experiments but its operation is challenging, due to the large electronic spin of the atoms. 
Here we present an experimental study of narrow-line Dysprosium MOTs. We show that the combination of radiation pressure and gravitational forces leads to a spontaneous polarization of the electronic spin. The spin composition is measured using a Stern-Gerlach separation of spin levels, revealing that the gas becomes almost fully spin-polarized for large laser frequency detunings. In this regime, we reach the optimal operation of the MOT,  with samples of typically $3\times 10^8$ atoms at a temperature of 15\,$\mu$K. The spin polarization reduces the complexity of the radiative cooling description, which allows for a simple model accounting for our measurements. We also measure the rate of density-dependent atom losses, finding good agreement with a model based on light-induced Van der Waals forces. A minimal two-body  loss rate  $\beta\sim 2\times10^{-11}\,$cm$^{3}$/s is reached in the spin-polarized regime. Our results constitute a  benchmark for the experimental study of ultracold gases of magnetic Lanthanide atoms.

\end{abstract}

\maketitle

\section{Introduction}

Open-shell Lanthanide atoms bring up new perspectives in the field of ultracold quantum gases, based on their unique physical properties. Their giant magnetic moment allows exploring the behavior of long-range interacting dipolar systems beyond  previously accessible regimes \cite{lu2011strongly,aikawa2012bose,lu2012quantum,aikawa2014observation,baier2016extended,ferrier2016observation,kadau2016observing}. The large electronic spin also leads to complex low-energy scattering between atoms, exhibiting chaotic behavior \cite{frisch2014quantum,maier2015emergence}. The atomic spectrum, which includes narrow optical transitions, further permits the efficient production of artificial gauge fields \cite{cui2013synthetic,nascimbene2013realizing,burdick2016long}.

This exciting panorama triggered the implementation of laser cooling techniques for magnetic Lanthanide atoms,  including Dysprosium \cite{youn2010dysprosium,lu2010trapping,maier2014narrow}, Holmium \cite{hemmerling2014buffer,miao2014magneto}, Erbium \cite{mcclelland2006laser,berglund2008narrow,frisch2012narrow,hemmerling2014buffer} and Thulium \cite{sukachev2010magneto,hemmerling2014buffer}.
Among these techniques, magneto-optical trapping using a narrow optical transition (linewidth $\Gamma\sim2\pi\times100\,$kHz) provides an efficient method to prepare atomic samples of typically $10^8$ Lanthanide atoms in the 10\,$\mu$K temperature range \cite{frisch2012narrow,maier2014narrow}, in analogy with the trapping of Sr and Yb atoms using the intercombination line \cite{kuwamoto1999magneto,katori1999magneto}. 

For two-electron atoms like Ca, Sr and Yb \cite{kuwamoto1999magneto,katori1999magneto,binnewies2001doppler}, the absence of electronic spin simplifies the operation and theoretical understanding of the MOT. On the contrary, the large value of the electron spin for Lanthanide atoms greatly complicates the atom dynamics, the modeling of which a priori requires accounting for optical pumping effects between numerous spin levels. 
Previous works on narrow-line MOTs with Lanthanide atoms mentioned a spontaneous spin polarization of the atomic sample  \cite{berglund2008narrow,lu2011strongly,aikawa2012bose}, but its quantitative analysis  was not explicitely described. 

In this article, we present a study of Dy magneto-optical traps operated on the 626-nm optical transition (linewidth $\Gamma=2\pi\times 136\,$kHz \cite{gustavsson1979lifetime}) \cite{maier2014narrow}. We measure the spin populations using a Stern-Gerlach separation of the spin levels. We observe that, for large and negative laser detunings (laser frequency on the red of the optical transition), the atomic sample becomes spin-polarized in the absolute ground state $\ket{J=8,m_J=-8}$. This spontaneous polarization occurs due to the effect of gravity, which pushes the atoms to a region with a relatively large magnetic field (on the order of 1\,G), leading to efficient optical pumping \cite{berglund2008narrow,lu2011strongly,aikawa2012bose}.   In the spin-polarized regime, the system can be simply described with a two-level atom model, in close relation with previous works on Sr magneto-optical traps \cite{loftus2004narrow} and narrow-line MOTs of magnetic Lanthanides \cite{berglund2008narrow,lu2011strongly,aikawa2012bose}. 

We show that the spin-polarized regime corresponds to optimal operating parameters for the magneto-optical trap, leading to  samples with up to  $N\simeq3\times10^8$ atoms and temperatures down to $T\simeq15$\,$\mu$K. This observation is supported by a study of density-dependent atom losses triggered by 
light-induced Van der Waals interactions between atoms.
 We observe that minimal loss rates are reached in the spin-polarized regime, as predicted by a simple model of atom dynamics in attractive molecular states.

\section{Preparation of magneto-optically trapped Dysprosium gases}

\begin{figure}
\includegraphics{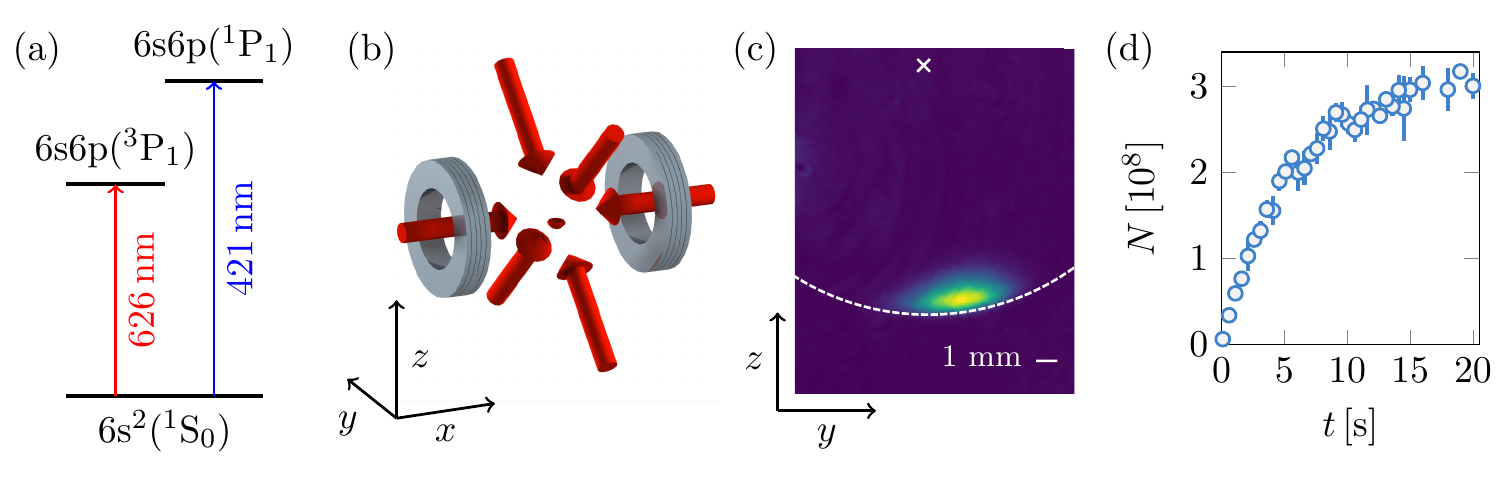}
\caption{
(a) Scheme of the optical transitions involved in our laser cooling setup, coupling the electronic ground state 4f$^{10}$($^5$I$_8$)6s$^2$($^1$S$_0$) ($J=8$)  to the excited states 4f$^{10}$($^5$I$_8$)6s6p($^1$P$_1$)(8,1)$_9$ ($J'=9$) and 4f$^{10}$($^5$I$_8$)6s6p($^3$P$_1$)(8,1)$_9$ ($J'=9$).
(b) Scheme of the magneto-optical trap arrangement, with the six MOT beams (red arrows) and the pair of coils in anti-Helmoltz configuration. Gravity is oriented along $-z$.
 (c) Typical in situ absorption image of an atomic sample held in the `compressed' magneto-optical trap. The cross indicates the location of the quadrupole center. The cloud is shifted by $\sim1\,$cm below the zero due to gravity. We attribute the cloud asymmetry with respect to the $z$ axis to the effect of an ambient magnetic field gradient. 
(d) Atom number $N$ captured in the magneto-optical trap  as a function of the loading time $t$ (see text for the MOT parameters). 
\label{Fig_Scheme}}
\end{figure}

The electronic states of Dysprosium involved in our study  are represented in figure\,\ref{Fig_Scheme}a, together with a schematics of the magneto-optical trap in figure\,\ref{Fig_Scheme}b. 
An atomic beam is emitted by an effusion cell oven, heated up to 1100\,°C.
The $^{164}$Dy atoms are decelerated in a Zeeman slower, which is built in a spin-flip configuration and operates on the broad optical transition at 421\,nm (linewidth $2\pi\times32\,$MHz \cite{lu2011spectroscopy}). The flux of atoms along the Zeeman slower axis is enhanced using transverse Doppler cooling at the entrance of the Zeeman slower, also performed on the 421-nm resonance \cite{leefer2010transverse}. 

The magneto-optical trap, loaded in the center of a steel chamber, uses a quadrupole magnetic field of gradient $G=1.71$\,G/cm along the strong horizontal axis $x$. 
The MOT beams, oriented as pictured in figure \ref{Fig_Scheme}b, operate at a frequency on the red of the optical transition at 626\,nm \cite{maier2014narrow}, the detuning from resonance being denoted $\Delta$ hereafter. This transition connects the electronic ground state 4f\textsuperscript{10}6s\textsuperscript{2}\,($^5$I$_8$), of angular momentum $J=8$ and Landé factor $g_J\simeq1.24$, to the electronic state 4f\textsuperscript{10}($^5$I$_8$)\,6s\,6p($^3$P$_1$)\,(8,1)$_9$, of angular momentum $J'=9$ and Landé factor $g_{J'}\simeq1.29$. Its linewidth $\Gamma=2\pi\times136\,$kHz, corresponding to a saturation intensity $I_{\mathrm{sat}}=72\,\mu$W/cm$^2$, allows in principle for Doppler cooling down to the temperature $T_\mathrm{D}=\hbar\Gamma/(2k_B)\simeq3.3\,\mu$K.
Each MOT beam is prepared with  a waist $w\simeq20$ mm and an intensity  $I=3.7\,$mW/cm$^2$ on the beam axis, corresponding to a saturation parameter $s\equiv I/I_{\mathrm{sat}}\simeq 50$. 

The atom loading rate is increased by artificially broadening the MOT beam frequency: The laser frequency is sinusoidally modulated  at  135\,kHz, over a total frequency range of 6\,MHz, and with a mean laser detuning $\Delta = -2\pi\times4.2$\,MHz. From a typical atom loading curve (see  figure\,\ref{Fig_Scheme}d) one obtains a loading rate of $6(1)\times10^7$ atoms/s  at short times and a maximum atom number $N=3.1(5)\times10^8$. 
The atom number is determined up  to a 20\% systematic error using absorption imaging with resonant light on the broad optical transition, taking into account the variation of scattering cross-sections among the spin manifold expected for our imaging setup.

After a loading duration of $\sim 6\,$s, we switch off the magnetic field of the Zeeman slower, as well as the slowing laser. We then compress the magneto-optical trap by ramping down the frequency modulation, followed by  decreasing the saturation parameter $s$ and the laser detuning $\Delta$ over a total duration of 430\,ms. At the same time, the magnetic field gradient is  ramped to its final value. For most of the MOT configurations used for this study, we do not observe significant atomic loss during the compression.
A typical absorption image of the atomic sample after compression is shown in figure\,\ref{Fig_Scheme}c. The curved shape of the gas and its mean position (about 1\,cm below the magnetic field zero) reveal the role of gravity in the magneto-optical trapping \cite{katori1999magneto,loftus2004narrow,berglund2008narrow}. 

\section{Spin composition\label{section_spin_composition}}

\begin{figure}
\includegraphics{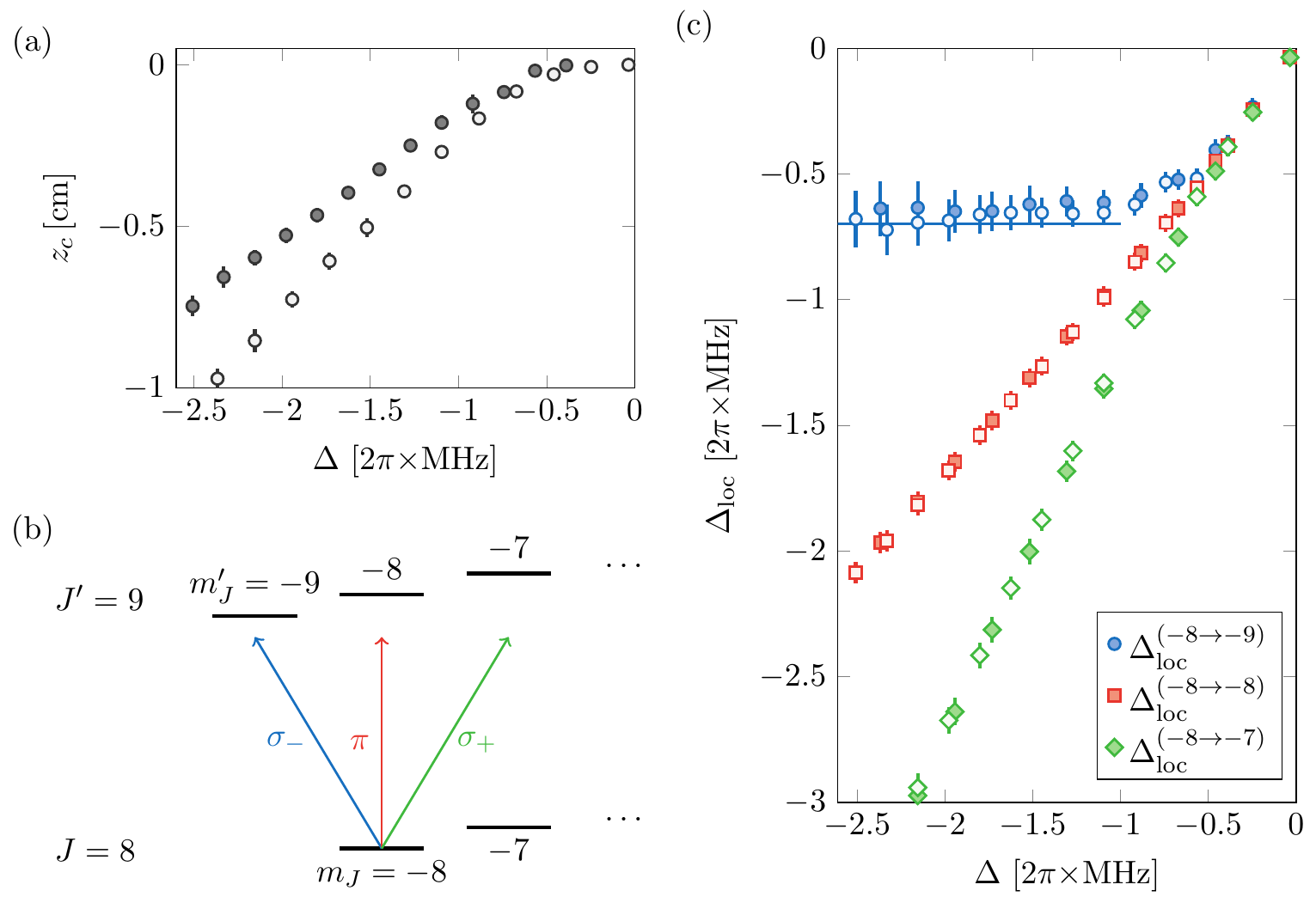}
\caption{
(a) Vertical position $z_c$ of the MOT center of mass as a function of the laser detuning $\Delta$, measured for two values of the magnetic field gradient, $G=1.71$\,G/cm (open symbols) and $G=2.26$\,G/cm (filled symbols) and a saturation parameter $s=0.65$.
 (b) Scheme of the optical transitions starting from the absolute ground state $\ket{J=8,m_J=-8}$. As the MOT position deviates from the magnetic field zero position, optical transitions of polarization $\sigma_-$ become more resonant than $\pi$ and $\sigma_+$ transitions, leading to optical pumping into the absolute ground state.
(c) Local detunings for the three optical transitions involving the absolute ground state, inferred from the laser detuning and the Zeeman shifts at the MOT position. We take into account an ambient magnetic field gradient along $z$, $\delta G=-0.094(2)$\,G/cm, measured independently. The local detuning of the $\sigma_-$ transition (circles) saturates to a fixed value for large detunings, showing that the MOT position adapts to keep the local detuning fixed. The $\pi$ and $\sigma_+$ transitions (square and diamonds, respectively) do not exhibit this saturation.
\label{Fig_Position}
}
\end{figure}

An immediate striking difference of narrow-line MOTs with respect to alkali-metal ones is the strong dependence of the MOT center position on detuning \cite{loftus2004narrow}. As shown in figure\,\ref{Fig_Position}a, we indeed observe a drop of the MOT position when increasing the laser detuning, the amplitude of which largely exceeds the cloud size (see figure \ref{Fig_Scheme}c). This behavior can be explained by considering the MOT equilibrium condition, which requires mean radiative forces to compensate for gravity. When the laser detuning is increased, the MOT position adapts  so as to keep the mean amplitude of radiative forces constant.

This picture is supported by the calculation of the \emph{local} detunings $\Delta_{\mathrm{loc}}^{(m_J\rightarrow m_J')}$ of optical transitions $\ket{J=8,m_J}\rightarrow \ket{J'=9,m_J'}$ at the MOT position.  We compare in figure\,\ref{Fig_Position}c these detuning values for the $\sigma_-$, $\pi$ and $\sigma_+$ transitions starting from the ground state $\ket{J=8,m_J=-8}$. We observe that, when increasing the detuning $\Delta$, the $\pi$ and $\sigma_+$ transitions become off-resonant, while the local detuning of the $\sigma_-$ transition tends to a finite value. The detuning of the latter transition is denoted in the following as $\Delta_{\mathrm{loc}}\equiv\Delta_{\mathrm{loc}}^{(-8\rightarrow -9)}$.

\begin{figure}
\includegraphics{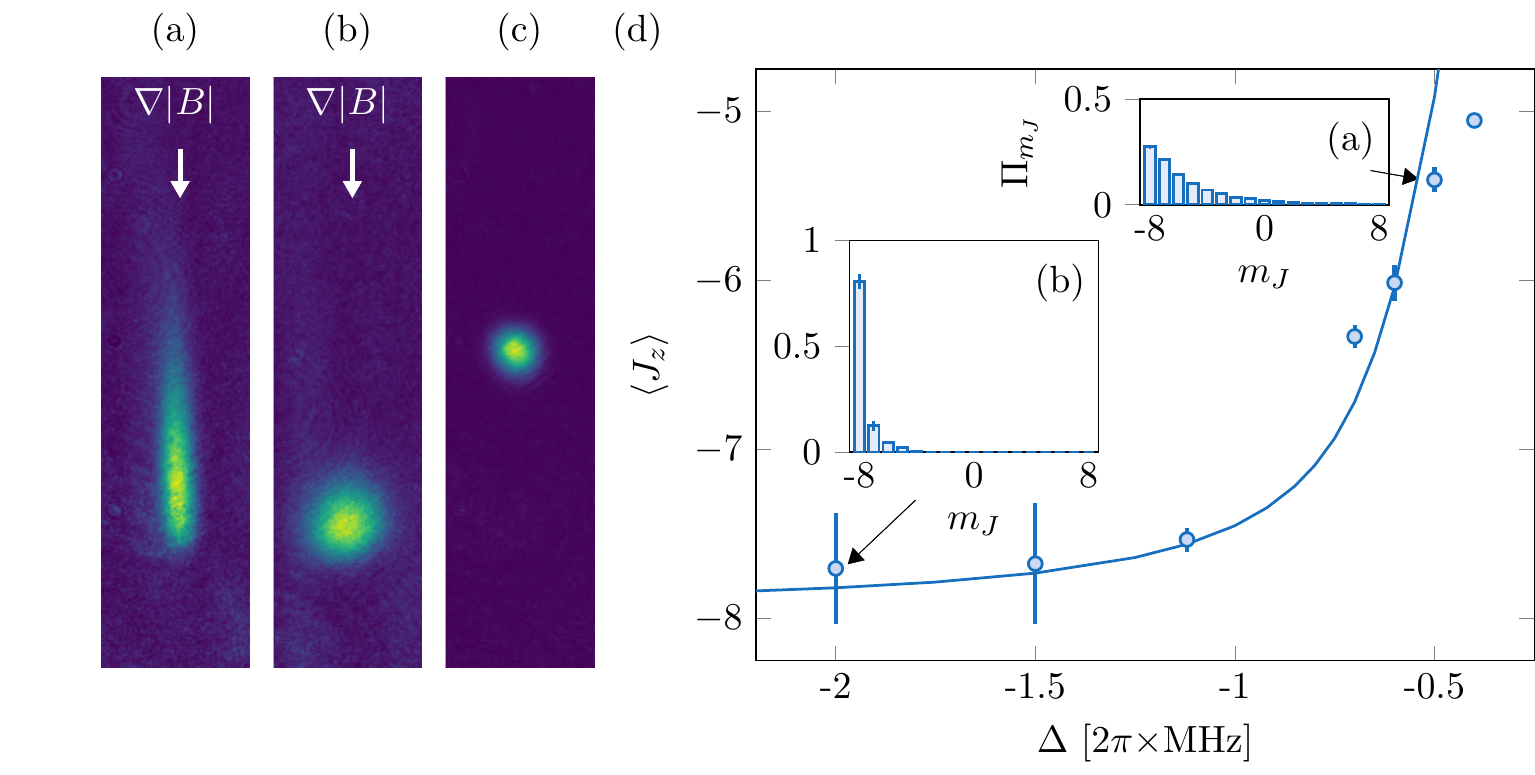}
\caption{(a), (b) Typical absorption images of the atomic samples taken in our Stern-Gerlach experiment (see text). The populations of individual spin levels can be resolved using a multiple gaussian fit of the atom density.  Image (c) serves as a reference image taken in the absence of gradient. (d)  Variation of the mean spin projection  $\langle J_z\rangle$ with the laser detuning $\Delta$, showing the almost full polarization in the absolute ground state, i.e. $\langle J_z\rangle\simeq -8$, for large detunings ($\Delta\lesssim-2\pi\times 1$\,MHz). The solid line corresponds to the prediction of an optical pumping model (see text). Insets show the measured spin population distributions $\Pi_{m_J}$ for the detunings $\Delta/(2\pi)=-0.5$\,MHz  and $-2$\,MHz, corresponding to images (a) and (b), respectively.  The saturation parameter is $s=0.65$ and the MOT gradient is $G=1.71\,$G/cm. 
\label{Fig_Stern_Gerlach}}
\end{figure}

This predominance of the $\sigma_-$ transition leads to  optical pumping of the electronic spin towards the absolute ground state, as confirmed by a direct measurement of the spin composition of the atomic sample with a Stern-Gerlach separation of spin levels. To achieve this, we release the atoms from the MOT, apply a vertical magnetic field gradient of about 30\,G/cm during $\sim 4\,$ms, and let the atoms expand for 20\,ms before taking an absorption picture (see figures\;\ref{Fig_Stern_Gerlach}a,b,c). 
We measure the spin composition for several detuning values $\Delta$, and we show two examples of spin population distributions in the insets of figure\,\ref{Fig_Stern_Gerlach}d. The mean spin projection $\langle J_z \rangle$ inferred from these data is plotted as a function of the detuning $\Delta$ in figure\,\ref{Fig_Stern_Gerlach}d. It reveals an almost full polarization for large detunings $\Delta\lesssim -2\pi\times 1\,$MHz, which is the first main result of our study.

In the spin-polarized regime, the theoretical description of the magneto-optical trap can be simplified. As the atomic gas is spin-polarized and the $\sigma_-$ component of the MOT light dominates over other polarizations, the atom electronic states can be restricted to a two-level system, with a ground state $\ket{J=8,m_J=-8}$ and an excited state $\ket{J'=9,m_J'=-9}$ \cite{loftus2004narrow}. The radiative force is then calculated by summing the contributions of the six MOT beams, projected on the $\sigma_-$ polarization. For the sake of simplicity we restrict here the discussion to the motion on the $z$ axis, but extending the model to describe motion along $x$ and $y$ directions is straightforward (see \ref{Appendix_Temperature_3D}). The motion along $z$ is governed by the radiative forces induced by the four MOT beams propagating in the plane $x=0$ (see figure \ref{Fig_Scheme}b). Taking only the $\sigma_-$ component of these beams into account  and neglecting interference effect between the beams, the total  force for an atom at rest on the $z$ axis is obtained by summing the radiation pressure forces from the four beams in the $x=0$ plane. It takes the simple form 
\[
\mathbf{F}_{\mathrm{rad}}=\frac{\hbar k \Gamma}{2}\frac{s}{1+2s+4\Delta_{\mathrm{loc}}^2/\Gamma^2}\mathbf{e}_z
\]
at the MOT position \cite{metcalf2007laser} \footnote{For a different MOT beam geometrical  configuration, with propagation directions along $\pm\mathbf{e}_x$, $\pm\mathbf{e}_y$ and $\pm\mathbf{e}_z$, we would obtain the same expressions for the radiative force, as well as the optical pumping rates described in the following.}.
The MOT equilibrium position corresponds to the condition of the radiative force compensating gravity $\mathbf{F}_{\mathrm{rad}}+m\mathbf{g}=\boldsymbol 0$, leading to
\begin{equation}\label{eq_equilibrium}
\frac{s}{1+2s+4\Delta_{\mathrm{loc}}^2/\Gamma^2}=\frac{1}{\eta},\quad\eta=\frac{\hbar k \Gamma}{2mg},
\end{equation}
where $\eta\simeq168$ for the considered optical transition.
The local detuning $\Delta_{\mathrm{loc}}$ thus only depends on the laser intensity $s$ and not on the bare detuning $\Delta$, as
\begin{equation} 
\label{eq_Dloc}
\Delta_{\mathrm{loc}}=-\frac{\Gamma}{2}\sqrt{(\eta-2)s-1}.
\end{equation}
This expression accounts well for the experimental data presented in figure\,\ref{Fig_Position}c: the saturation of the local detuning $\Delta_{\mathrm{loc}}$ for large detunings corresponds to the spin-polarized regime, where equation (\ref{eq_Dloc}) applies. For simplicity we do not take into account magnetic forces associated with the magnetic field gradient, as it leads to $\sim$$ 10\%$ corrections on the local detuning  $\Delta_{\mathrm{loc}}$, which is below our experimental error bars.

To go further this simple approach, we  also developed a model taking into account the populations in all Zeeman sublevels. The  Zeeman populations in the ground and excited states are calculated as the stationary state of optical pumping rate equations, including the Zeeman shifts corresponding to  a given position. We then calculate the radiative force by summing the contributions of all optical transitions, which allows determining the MOT position $z_c$ from the requirement of compensation of the radiative force and gravity.  As shown in figure\;\ref{Fig_Stern_Gerlach}d, this model accounts well for the measured population distributions.

\section{Equilibrium temperature}
\begin{figure}
\label{Fig_Temperature}
\includegraphics{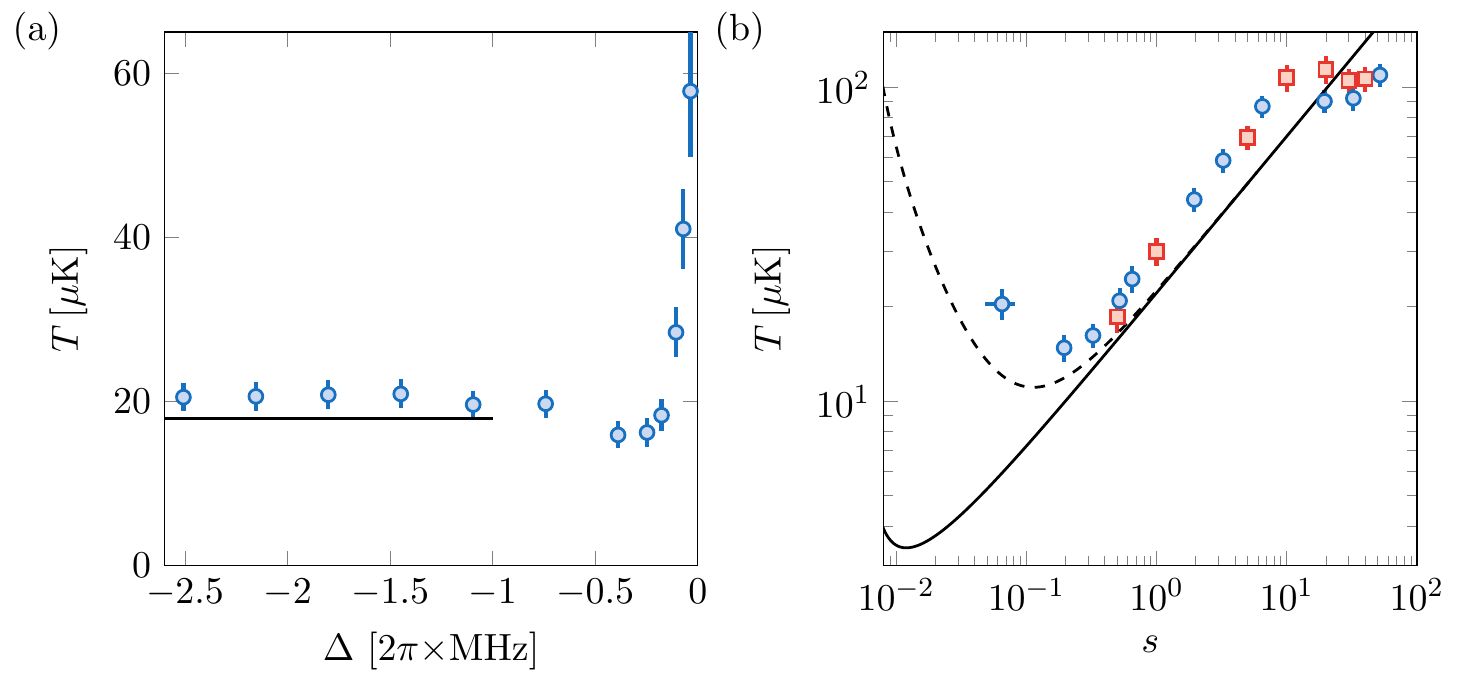}
\caption{
(a) Temperature $T$ of the atomic gas as a function of the laser detuning $\Delta$, for a saturation parameter $s=0.65$ and a gradient value $G=1.71\,$G/cm. 
(b) Temperature $T$ as a function of the saturation parameter $s$, measured for a gradient $G=1.71\,$G/cm and laser detunings $\Delta=-2\pi\times1.84\,$MHz (blue dots) and $\Delta=-2\pi\times2.54\,$MHz (red squares). The solid  lines in (a) and (b) correspond to the temperature expected in the spin-polarized regime according to equation (\ref{eq_T}). The dashed line includes the temperature increase expected from the measured intensity noise of the cooling laser beams (see \ref{Appendix_noise}).
}
\end{figure}

The main interest in using narrow optical transitions for magneto-optical trapping lies in the low equilibrium temperatures, typically in the 20\,$\mu$K range. In order to investigate the effect of the spin composition of the gas on its equilibrium temperature, we investigated the variation of the temperature $T$ -- measured after time-of-flight\footnote{The magnetic field gradient is kept on during expansion in order to avoid eddy current effects. We checked that magnetic forces play a negligible role in the expansion dynamics for the flight durations used for this measurement.} --  with the laser detuning $\Delta$ (see figure\,\ref{Fig_Temperature}). 
Far from resonance, we observe that  the temperature does not depend on $\Delta$, in agreement with the picture of the spin-polarized regime discussed above (see figure\,\ref{Fig_Temperature}a). The temperature slightly decreases for $-1\,\mathrm{MHz}<\Delta/(2\pi)<-0.3\,$MHz, i.e. when leaving the spin-polarized regime, before significantly raising closer to resonance. 

We also investigated the influence of the laser intensity, by probing the temperature variation with the saturation parameter $s$ (see figure\,\ref{Fig_Temperature}b). The observed temperature raise upon increasing $s$  can be intuitively understood from equation\,(\ref{eq_Dloc}): the local detuning from resonance increases when raising $s$; we then expect a temperature increase according to the Doppler cooling theory (in the regime $|\Delta_{\mathrm{loc}}|>\Gamma/2$ considered here).

A more quantitative understanding requires adapting the theory of Doppler cooling to the experiment geometry.  Here we restrict the discussion to the atom dynamics along $z$, in  the spin-polarized regime (see \ref{Appendix_Temperature_3D} for a generalization to 3D). The radiative force produced by the four MOT beams in the $x=0$ plane can be calculated for an atom of velocity $v$ along $z$, leading to the damping force for small velocities
\begin{equation}
\label{eq_alpha}
F_{\mathrm{rad}}=-m\alpha v,	\quad	\alpha=-3\frac{\hbar k^2}{m}\frac{s\,\Delta_{\mathrm{loc}}/\Gamma}{[1+2s+4(\Delta_{\mathrm{loc}}/\Gamma)^2]^2},
\end{equation}
which coincides with the usual damping force formula for Doppler cooling for a pair of counter-propagating laser beams of saturation parameter $s$ and detuning $\Delta_{\mathrm{loc}}$, up to numerical factors related to the geometry of our laser configuration.  In \ref{appendix_temp}, we discuss additional experiments on temperature equilibration dynamics, which can be explained qualitatively using the damping rate value (\ref{eq_alpha}).

The momentum diffusion coefficient along $z$, denoted as $D_{zz}$, is calculated taking into account the contribution of the six cooling laser beams \cite{wineland1979laser}, as
\[
D_{zz}=\frac{31}{80}\hbar^2k^2\Gamma\,\Pi',
\]
where $\Pi'={1}/{\eta}$ is the population in the excited state. The temperature $T$ is then obtained as the ratio $k_BT=D_{zz}/(m\alpha)$, leading to the expression
\begin{equation}\label{eq_T}
T=\frac{31}{120}\frac{\eta s}{\sqrt{s(\eta-2)-1}}\frac{\hbar\Gamma}{k_B}.
\end{equation}
Extending this analysis of the atom dynamics to the two other spatial directions $x$ and $y$ leads to slightly different equilibrium temperatures along these axes, but the expected difference is  less than 10\%, which is below our experimental resolution (see \ref{Appendix_Temperature_3D}).
According to equation (\ref{eq_T}) , the MOT temperature is minimized down to 
\[
T_{\rm{min}}=\frac{31}{60}\frac{\eta}{(\eta-2)}\frac{\hbar\Gamma}{k_B}\simeq3.4\,\mu\mathrm{K}
\]
for $s=2/(\eta-2)$, which corresponds to a local detuning $\Delta_{\mathrm{loc}}=\Gamma/2$ according to equation (\ref{eq_Dloc}). As $\eta\gg1$, this value is very close to the standard Doppler limit $T_{\mathrm{D}}=\hbar\Gamma/(2k_B)$. 

We investigated this behavior by measuring the gas temperature as a function of the saturation parameter $s$ for large laser detunings. As shown in figure\,\ref{Fig_Temperature}b, the measured temperatures are reasonably well reproduced by equation\;(\ref{eq_T}) for $0.3\leq s\leq 10$. The measured temperatures deviate from theory for small saturation parameter values $s\lesssim0.1$, which can be explained by the shaking of the atomic sample due to the noise of the  cooling laser intensity. We calculate in \ref{Appendix_noise} the temperature increase expected from the measured intensity noise spectrum  (corresponding to  r.m.s. fluctuations of the saturation parameter $s$ of $5\times10^{-3}$). We obtain the dashed curve in  figure \ref{Fig_Temperature}b, which qualitatively reproduces the  temperature raise observed for small saturation parameters. A minimal temperature of 15(1)\,$\mu$K is measured for $s=0.2$.
For large saturation parameters $s\gtrsim 10$, the gas does not remain spin-polarized, and the temperature is no longer accounted for by equation\,(\ref{eq_T}). 

We also observe a raise of the MOT temperature when increasing the atom number, similarly to previous studies on alkali and alkaline-earth atoms \cite{drewsen1994investigation,boiron1996laser,katori1999magneto,kerman2000beyond}. We discuss this effect in the \ref{appendix_temp}. 

\section{Cloud sizes and atom density\label{section_geometry}}
\begin{figure}
\includegraphics{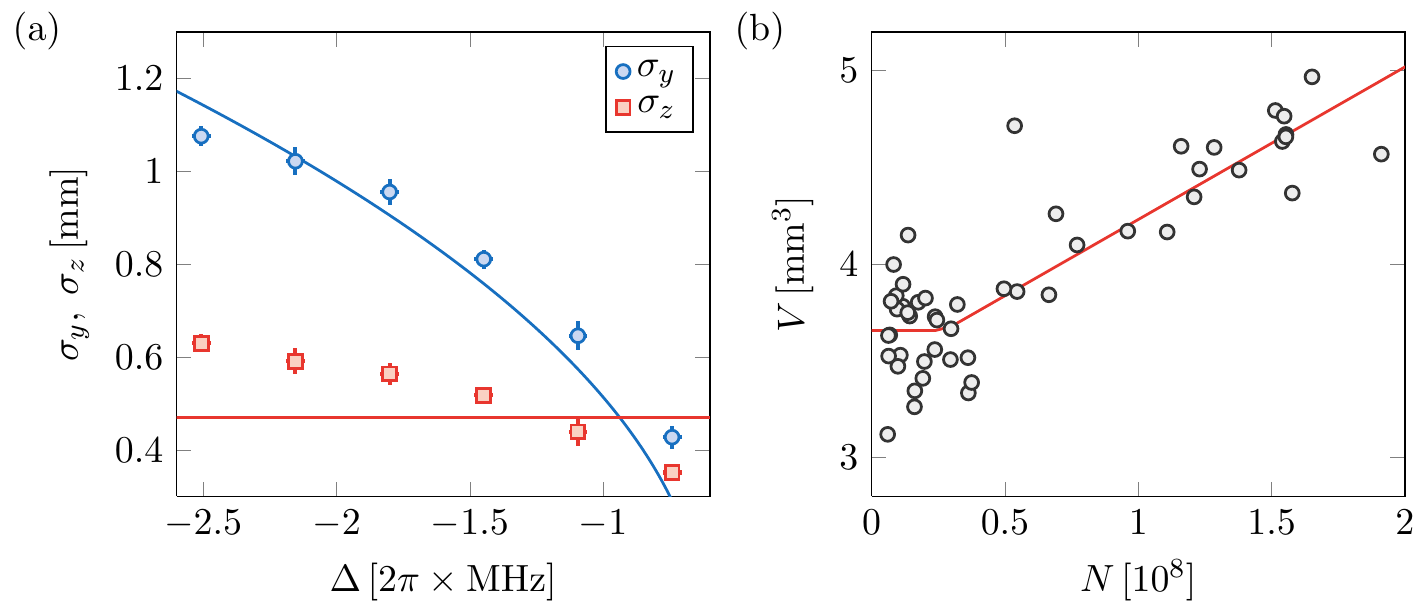}
\caption{
(a) R.m.s. cloud sizes $\sigma_y$ and $\sigma_z$  (blue dots and red squares, respectively), measured using absorption images taken in situ for $s=0.65$, $G=1.71\,$G/cm and $N\sim 2\times 10^7$. The solid lines correspond to the theoretical values expected in the spin-polarized regime, equations\,(\ref{eq_omega_x}) and (\ref{eq_omega_z}).
(b) Volume $V$ of the atomic gas as a function of the atom number $N$, for $s=0.65$, $\Delta=-2\pi\times 1.84\,$MHz and $G=1.71\,$G/cm. The solid line is a piecewise linear fit to the experimental data (see equation \ref{eq_V}), consistent with a maximum density reachable in our MOT $n_{\mathrm{max}}=7(1)\times 10^{10}\,\mathrm{cm}^{-3}$.

}
\label{Fig_Frequencies}
\end{figure}

The atom density in the MOT is an important parameter to consider for efficiently loading the atoms into an optical dipole trap. In this section we characterize the cloud sizes and atom densities  achieved in our setup.  We measured the horizontal and vertical r.m.s. sizes $\sigma_y$ and $\sigma_z$, respectively, using a gaussian fit of the optical density -- the absorption image being taken in situ (see figure\,\ref{Fig_Frequencies}a). We observe that, in the spin-polarized regime, the vertical size $\sigma_z$ weakly varies upon an increase of the laser detuning $\Delta$, while the horizontal size $\sigma_y$ increases over a larger range.

We now explain how one can account for this behavior within the simple model developed above. In the spin-polarized regime, the analytic form of the radiative force allows expressing the equilibrium shape of the atomic sample in a simple manner. Close to the equilibrium position, the radiative force can be expanded linearly as $\mathbf{F}_{\mathrm{rad}}=-\kappa_x x\,\ex-\kappa_y y\,\ey-\kappa_z z\,\ez$, where the spring constants are given by 
\begin{eqnarray}
\kappa_x&=\frac{2mg}{|z_c|}\label{eq_omega_x},\\
\kappa_y&=\kappa_x/2\label{eq_omega_y},\\
\kappa_z&=\frac{4mg\,\delta\mu\, G|\Delta_{\mathrm{loc}}|}{s\,\eta\,\hbar\,\Gamma^2}\label{eq_omega_z},
\end{eqnarray}
where $\delta\mu=\mu'-\mu$ is the difference between the magnetic moments in the excited and ground electronic states, denoted as $\mu'$ and $\mu$, respectively. Note the simple expression (\ref{eq_omega_x}) for $\kappa_x$, in which the influence of the detuning $\Delta$ and saturation parameter $s$ only occurs via the equilibrium position $z_c$.  We remind that the magnetic field gradient is twice larger along $x$ than along $y$, which explains the relation (\ref{eq_omega_y}) between the spring constants $\kappa_x$ and $\kappa_y$. The r.m.s. cloud sizes $\sigma_u$ ($u=x,y,z$) are then determined using the thermodynamic relations 
\begin{equation}\label{eq_sizes}
k_B T=\kappa_u\sigma_u^2.
\end{equation}
As both the temperature $T$ and local detuning $\Delta_{\mathrm{loc}}$ are constant in the spin-polarized regime, equations (\ref{eq_omega_z}) and (\ref{eq_sizes}) predict a constant r.m.s. size $\sigma_z$, consistent with the weak variation observed experimentally. The variation of the size $\sigma_y$ is also well captured by this model. A more precise analysis would require taking into account trap anharmonicities, which cannot be completely neglected given the non-gaussian cloud shape (see figure \ref{Fig_Scheme}c).

We also observe a variation of the cloud size when increasing the atom number $N$ with fixed MOT parameters (see figure\;\ref{Fig_Frequencies}b). Such an effect is expected from the repulsive interaction between atoms being dressed by the MOT light, due to the radiation pressure of fluorescence light they exert on each other \cite{walker1990collective}. We plotted in figure\,\ref{Fig_Frequencies}b the cloud volume $V$ as a function of the atom number $N$. The volume $V$ is defined as $V=(2\pi)^{3/2}\sigma_x\sigma_y\sigma_z$, so that $n_{\mathrm{peak}}=N/V$ represent the atom density at the trap center. In order to calculate the volume $V$, we use equation\,(\ref{eq_omega_y}) to deduce $\sigma_x$ from the measured $\sigma_y$ value. The measurements are consistent with a volume $V$ independent of $N$ for low atom numbers (`temperature-limited' regime) and linearly varying with $N$ for large atom numbers (`density-limited' regime) \cite{walker1990collective,townsend1995phase}. The data is fitted with the empirical formula 
\begin{equation}\label{eq_V}
V=V_{\mathrm{single\;atom}}+\alpha (N-N_c)\Theta(N-N_c),
\end{equation}
where $\Theta$ is the Heaviside function. For the MOT parameters corresponding to  figure \ref{Fig_Frequencies}b, we obtain $V_{\mathrm{single\;atom}}=3.7(1)\,$mm$^3$, $N_c=3(1) \,10^7$ and $\alpha=8(1)\times10^{-9}\,$mm$^3$.

Far in the density-limited regime ($N\gg N_c$), we expect the atoms to organize as an ellipsoid of uniform atom density $n_{\mathrm{max}}$ corresponding to the maximum atom density that can be reached in the MOT \cite{walker1990collective,townsend1995phase}. In such a picture, the volume determined from the r.m.s. sizes varies linearly with the atom number, with $\alpha=0.52\,n_{\mathrm{max}}^{-1}$ (taking into account the non-gaussian atom distribution in this regime). From the fit (\ref{eq_V}) of our data, we infer for large atom numbers a maximum atom density $n_{\mathrm{max}}=7(1) \,10^{10}\,\mathrm{cm}^{-3}$, a value comparable to the ones typically reached with alkali atoms \cite{townsend1995phase}.

\section{Atom losses due to light-assisted collisions\label{section_light-induced_collisions}}
The variation of the cloud sizes shown in figure \ref{Fig_Frequencies} indicates that the atom density could be maximized by setting the cooling laser light close to the optical resonance, so as to achieve the smallest cloud volume. However, we observe an increased rate of atom losses near resonance, eventually leading to a reduced atom density. In order to understand this behavior, we present in this section an experimental study of atom losses in the magneto-optical trap, and we interpret the measurements with a simple model  based on  molecular dynamics resulting from light-induced Van-der-Waals interactions \cite{gallagher1989exoergic,julienne1991cold,julienne1992theory,dinneen1999cold}.

\begin{figure}
\includegraphics{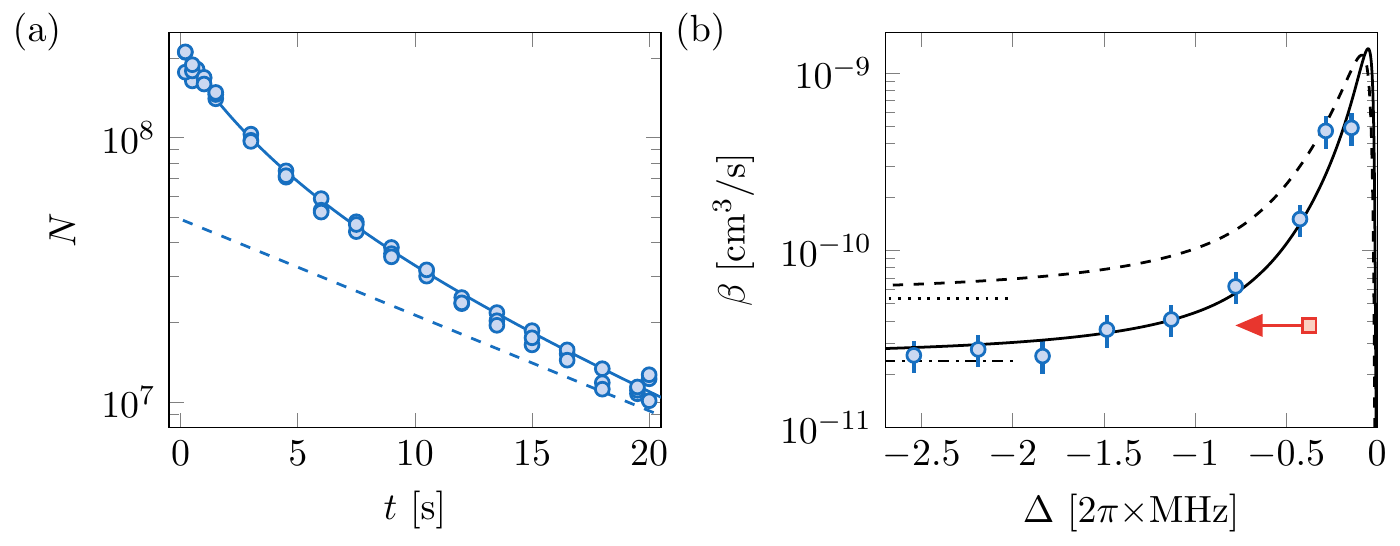}
\caption{
(a) Decay of the atom number with time  for a laser detuning $\Delta=-2\pi\times2.54$\,MHz, a saturation parameter $s=0.65$ and a gradient $G=1.71\,$G/cm. The data is fitted with a solution of the decay model (\ref{eq_loss_model}) (solid line), leading to a two-body loss coefficient $\beta=2.3(5)\times10^{-11}$\,cm$^3$/s. The dashed line corresponds to the exponential asymptote associated with one-body atom losses ($1/e$ time constant $\tau_1=12\,$s).
(b) Variation of the loss coefficient $\beta$ with the detuning $\Delta$, compared with the predictions of the molecular dynamics model. The dashed line corresponds to molecular parameters $\lambda = \bar{\lambda}$ and $\mu = \bar{\mu}$, while the solid line is a fit with $\lambda$ and $\mu$ being free parameters. The dotted and dash-dotted lines stand for the corresponding asymptotic expression (\ref{eq_beta_pol_simple}). The red square corresponds to the decay coefficient measured in reference \cite{maier2014narrow}, with  the arrow indicating the required detuning renormalization (see text).
}
\label{Fig_Atom_Losses}
\end{figure}

The loss of atoms is quantitatively characterized by measuring the variation of the atom number $N$ with the time $t$ spent in the magneto-optical trap, in the absence of Zeeman slowing light (see figure\,\ref{Fig_Atom_Losses}a). The atom decay is fitted with the solution of an atom loss model taking into account one-body atom losses due to collisions with the residual gas and two-body losses, described by the equation
\begin{equation}\label{eq_loss_model}
\dot N=-\frac{N}{\tau_1}-\beta\, \bar n N.
\end{equation}
In this equation, $\tau_1=12(2)\,$s is the one-body lifetime due to collisions with background atoms (background pressure $\simeq4\times10^{-10}\,$mbar), $\beta$ is the two-body loss coefficient and $\bar n=n_{\mathrm{peak}}/(2\sqrt{2})=N/(2\sqrt{2}V)$ is the average atom density in the trap. 
An example of fit of the atom number decay is presented in figure\,\ref{Fig_Atom_Losses}a\footnote{Before fitting the measured atom decay data with equation\,(\ref{eq_loss_model}), we fit the measured volume variation  with the atom number using the piecewise linear function (\ref{eq_V}), as discussed in section \ref{section_geometry}. }.

The figure \ref{Fig_Atom_Losses}b shows the variation of the loss coefficient $\beta$ with the detuning $\Delta$. We observe that the loss coefficient stays almost constant in the spin-polarized regime, with $\beta=2.6(5)\times10^{-11}$\,cm$^3$/s, and it increases by a factor $\sim 20$ when approaching  resonance. We can also compare our measurements to the loss coefficient $\beta=3.7(4)\times10^{-11}$\,cm$^3$/s reported in reference \cite{maier2014narrow}. As shown in figure \ref{Fig_Atom_Losses}b, the two measurements are in good agreement after renormalizing the laser detuning to account for the different saturation parameters used in the two studies, so as to compare atom samples with identical spin composition\footnote{Close to the spin-polarized regime, we expect the atom fraction in excited spin levels to scale as $s/\Delta^2$. In order to compare loss coefficients taken for a detuning $\Delta$ and saturation parameter $s'$ to data of saturation parameter $s$ with comparable spin composition, we thus renormalize the detuning as $\Delta\rightarrow\Delta\sqrt{s/s'}$. }. 

\begin{figure}
\begin{center}
\includegraphics{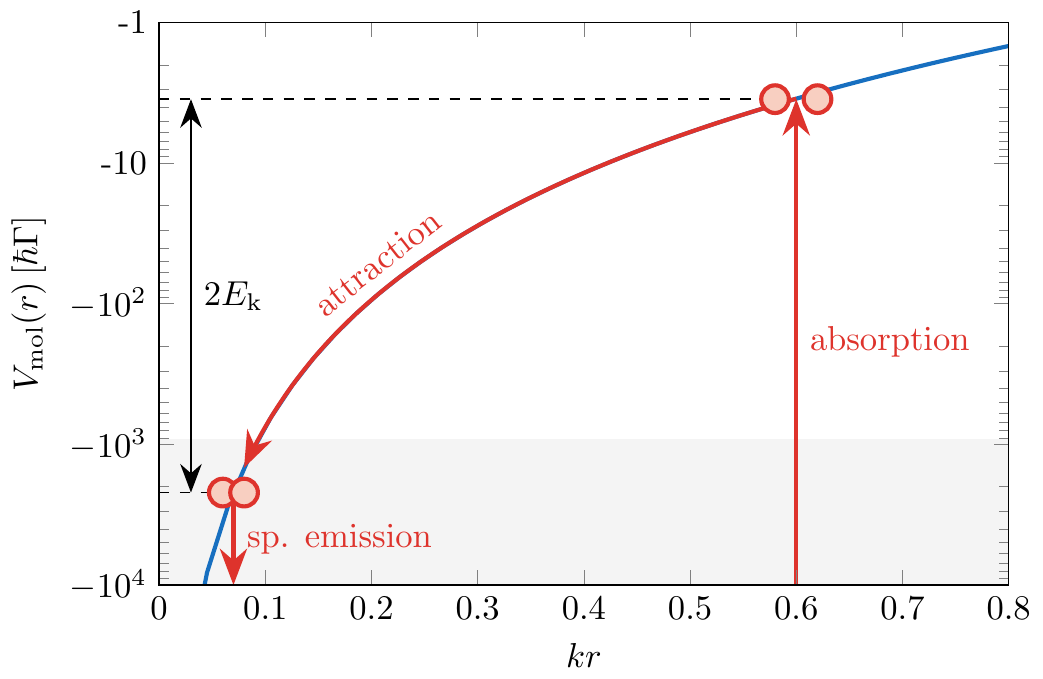}
\end{center}
\caption{Scheme of the light-induced inelastic collisions. For red-detuned laser light, atom pairs are preferentially excited to an attractive molecular potential (blue line). For the duration $\tau$ spent in the excited state, the atoms attract each other, and acquire each a kinetic energy $E_{\mathrm{k}}$ before returning to the electronic ground state. The atoms are lost as soon as their kinetic energy exceeds a threshold energy $E^*$, which is typically much larger than other involved energy scales (see \ref{Appendix_beta}). The condition for atom losses is represented as a light gray area, corresponding to $2\,E^*= 1000\,\hbar\Gamma$. }
\label{Fig_LI_collisions}
\end{figure}

We now interpret our loss coefficient measurements using a simple theoretical model in which atom losses originate from light-induced resonant Van der Waals interactions. 
As shown in figure\;\ref{Fig_LI_collisions}, when an atom is brought to an excited electronic state by absorbing one photon, it experiences strong Van der Waals forces from nearby atoms. For red-detuned laser light, an atom pair is preferentially promoted to attractive molecular potentials. Once excited in such a potential the pair rapidly shrinks and each atom may gain a large amount of kinetic energy. When the molecule spontaneously emits a photon, both atoms return to the electronic ground state with an additional kinetic energy that can be large enough for the atoms to escape the MOT.

We model this phenomenon using a simple description of molecular dynamics, inspired from \cite{gallagher1989exoergic,lett1995hyperfine,dinneen1999cold}. The large electron spin of Dysprosium leads to an intricate structure of $2(2J+1) (2J' +1) = 646$ molecular potential curves that we calculated numerically.  The complete description of this complex system is out of the scope of this paper. Fortunately, the main physical effects occurring in the experiment can be captured by a simplified model  corresponding to a single effective molecular potential $V_{\mathrm{mol}}(r)=-\lambda\hbar\Gamma/(kr)^3$, with a  $1/e$ molecule lifetime $(\mu\Gamma)^{-1}$, where $\lambda$ and $\mu$ are dimensionless numbers. The mean values of these parameters averaged  over the 323 attractive molecular potentials are $\bar\lambda\simeq0.68$ and $\bar\mu\simeq1.05$. 

The calculation of the two-body loss rate within this model is detailed in \ref{Appendix_beta}. We show that the laser excitation of atoms from the spin level $\ket{J,m_J}$ to $\ket{J',m_J+q}$  ($q=-1$, 0 or 1) contributes to the loss coefficient as 
\begin{eqnarray}\label{eq_beta}
\beta_{m_J,q}&=\Pi_{m_J}\frac{2\pi^2\lambda^2\mu}{3}\left|
\left<J',m_J+q\middle | J,m_J;1,q\right>
\right|^2\left(\frac{\Gamma}{\Delta_{\mathrm{loc}}^{(m_J\rightarrow m_J+q)}}\right)^2\frac{s\,\Gamma}{k^3}\nonumber\\
&\phantom{==} \times\exp\left[-C\left|\frac{\Gamma}{\Delta_{\mathrm{loc}}^{(m_J\rightarrow m_J+q)}}\right|^{5/6}\sqrt{\frac{\hbar\Gamma}{\Er}}\right],\\
 C&=\sqrt{\frac{\pi}{2}}\frac{\Gamma_{\mathrm{E}}(5/6)}{6\,\Gamma_{\mathrm{E}}(4/3)}\lambda^{1/3}\mu\simeq 0.264\,\lambda^{1/3}\mu,\nonumber
\end{eqnarray}
where $\Gamma_{\mathrm{E}}$ is the Euler Gamma function. We remind that $\Pi_{m_J}$ is the atom fraction in the state $\ket{J,m_J}$ and $\Delta_{\mathrm{loc}}^{(m_J\rightarrow m_J+q)}$ is the local detuning for the considered optical transition at the MOT position. The exponential factor corresponds to the probability that a molecular association event leads to the loss of the atom pair. The total loss coefficient $\beta$ is then obtained by summing the contributions $\beta_{m_J,q}$ of all optical transitions.

From equation (\ref{eq_beta}), we see that the atom losses associated with this mechanism are exponentially suppressed  when considering broad optical transitions. This suppression plays a large role for alkalis, for which other loss mechanisms dominate, e.g. fine-structure-changing collisions \cite{gallagher1989exoergic}. In the case considered here, the exponential factor has a moderate effect: for the data shown in figure \ref{Fig_Atom_Losses}b, it takes the value $\simeq 0.6$ for the transition $\ket{J,m_J=-J}\rightarrow\ket{J',m_J'=-J'}$ in the spin-polarized regime, assuming $\lambda=\bar\lambda$ and $\mu=\bar\mu$.

In the spin-polarized regime, the predominance of the transition $\ket{J,m_J=-J}\rightarrow\ket{J',m_J'=-J'}$ leads to a simpler expression for the loss coefficient 
\begin{equation}\label{eq_beta_pol_simple}
\beta=\beta_{-8,-1}=\frac{2\pi^2\lambda^2\mu}{3}\left(\frac{\Gamma}{\Delta_{\mathrm{loc}}}\right)^2\exp\left[-C\left|\frac{\Gamma}{\Delta_{\mathrm{loc}}}\right|^{5/6}\sqrt{\frac{\hbar\Gamma}{\Er}}\right]\frac{s\,\Gamma}{k^3}.
\end{equation}
As the local detuning $\Delta_{\mathrm{loc}}$ does not vary with $\Delta$ in this regime, we expect a constant loss coefficient $\beta$, as observed for the data presented in figure \ref{Fig_Atom_Losses}b for $\Delta\lesssim-2\pi\times 1\,$MHz. 

In figure \ref{Fig_Atom_Losses}b we show the prediction of the full model for two sets of values for the parameters $\lambda$ and $\mu$. Using the mean values $\bar\lambda$ and $\bar\mu$ only provides a qualitative description of the measured loss coefficients. A better agreement is obtained using $\lambda=0.75$ and $\mu=0.5$, possibly indicating the important role played by subradiant molecular states, which correspond to $\mu<1$.

\section{Conclusions and perspectives}
We presented a detailed experimental study of narrow-line magneto-optical trapping of Dysprosium, together with theoretical models supporting our measurements. We showed that the optimal operation of the MOT is obtained for large laser detunings, leading to a spontaneous spin polarization of the atomic sample and to minimal two-body atom losses. 

This understanding allows us to prepare gases in ideal conditions for transferring them into an optical dipole trap. In such a non-dissipative trap,  it is crucial to produce atomic samples polarized  in the electronic ground state in order to avoid dipolar relaxation \cite{
hensler2003dipolar,burdick2015fermionic}. In preliminary experiments, we were able to trap about $2\times10^{7}$  atoms into an optical dipole trap created by a single laser beam of wavelength $\lambda=1070\,$nm and optical power $P=40\,$W, focused to a waist of $35\,\mu$m. Optimal efficiency of the dipole trap loading is obtained by first preparing the MOT at the lowest  achieved  temperature of 15\,$\mu$K (corresponding to parameters $\Delta=-2\pi\times1.8\,$MHz, $s=0.2$ and $G=1.7\,$G/cm), and then superimposing   the dipole trap over the MOT center during 600\,ms. The phase space density of $\simeq 8\times10^{-5}$ reached after the dipole trap loading corresponds to a good starting point to reach quantum degeneracy via evaporative cooling \cite{lu2011strongly,lu2012quantum}. 

Our study will be of direct interest for magneto-optical traps of other atomic species featuring both narrow optical transitions and a spinful electronic state, such as the other magnetic Lanthanides. Future work could  investigate sub-Doppler cooling to very low temperatures in optical molasses. Contrary to sub-Doppler mechanisms observed with magnetic Lanthanides in broad-line magneto-optical traps \cite{berglund2007sub,youn2010dysprosium,youn2010anisotropic}, reaching temperatures below the Doppler limit of the 626 nm optical transition  would require applying an optical molasses  at low magnetic field \cite{dalibard1989laser,ungar1989optical}. 

\ack 

We thank E. Wallis and T. Tian for their contribution in the early stage of the experiment. This work is supported by the European Research Council
(Synergy grant UQUAM) and the Idex PSL Research University (ANR-10-IDEX-0001-02 PSL\PSLstar). L. S. acknowledges the support from the European Union (H2020-MSCA-IF-2014 grant n°661433).

\appendix

\settocdepth{section}

\section{Additional temperature measurements\label{appendix_temp}}
In this appendix we describe further temperature measurements related to the equilibration dynamics and to the influence of the atom density.

\begin{figure}
\includegraphics{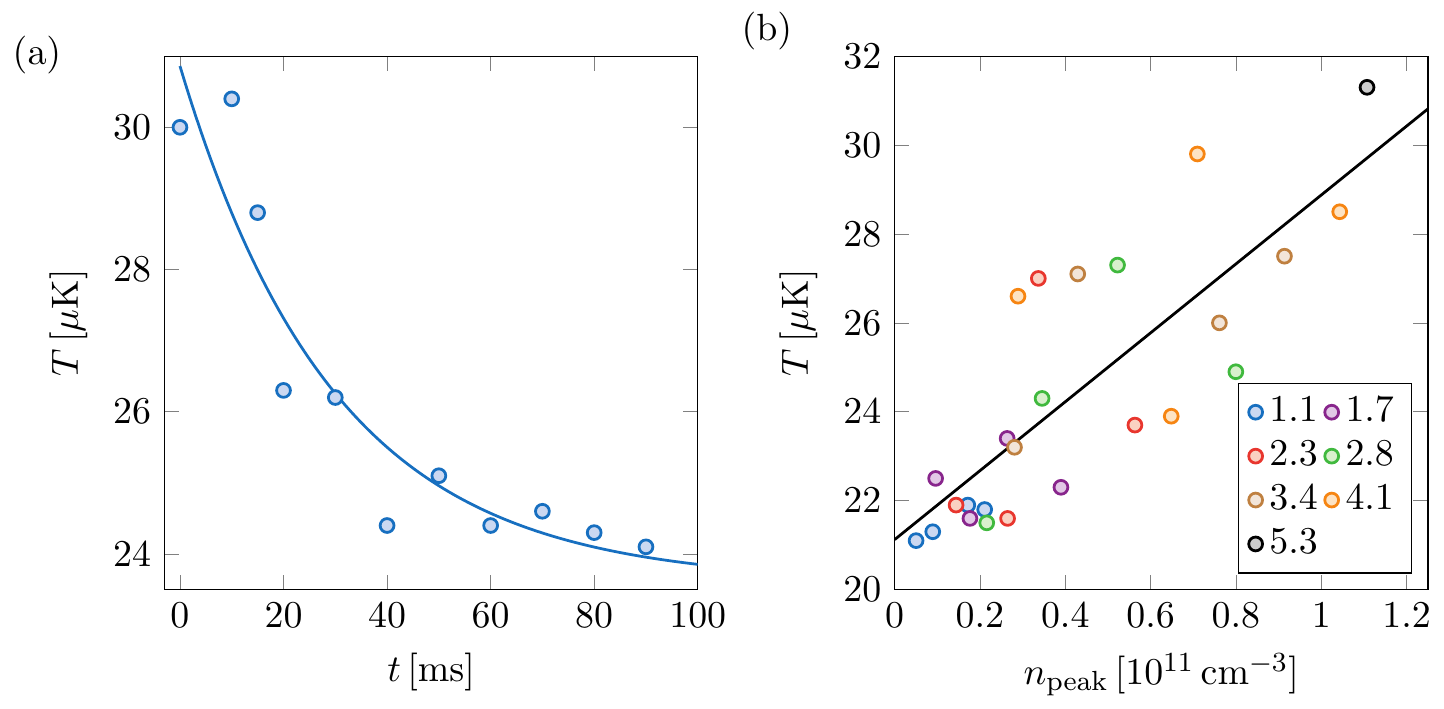}
\caption{
(a) Evolution of the gas temperature $T$ with time $t$ right after the MOT parameters are set to the `compressed MOT' values. The solid line is an exponential fit of the equilibration dynamics.
(b) Temperature $T$ as a function of atom density $n$, measured for various atom numbers and gradient values, indicated in the legend (in G/cm). The solid line is a linear fit of slope $\gamma=8(1)\times 10^{-11}\,\mu\mathrm{K}\,\mathrm{cm}^3$. 
\label{Fig_Temperature_Appendix}
}
\end{figure}

We studied the equilibration dynamics in the magneto-optical trap by measuring the time evolution of the temperature right after the trap parameters have been set to the `compressed MOT' values. We show such an evolution  in figure\,\ref{Fig_Temperature_Appendix}a, corresponding to MOT parameters $s=0.65$, $\Delta=-2\pi\times1.84$\,MHz and $G=1.71\,$G/cm. The temperature variation is fitted with an exponential decay of $1/\mathrm{e}$ time constant $\tau=29(11)\,$ms, with a baseline of $23.6(9)\,\mu$K. No significant atom loss is observed over the duration of equilibration.  This measurement allows extracting a damping coefficient $\alpha=1/(2\tau)=17(6)\,\rm{s}^{-1}$, comparable but smaller than the value $\alpha=47(2)\,\rm{s}^{-1}$ given by the simplified model leading to equation\;(\ref{eq_alpha}). 

We also investigated the raise of the MOT temperature when increasing the atom number \cite{drewsen1994investigation,boiron1996laser,katori1999magneto,kerman2000beyond}.
In previous studies, such an effect was attributed to multiple scattering of photons within the atomic sample \cite{boiron1996laser}, leading to a temperature raising  linearly with the peak atom density $n_{\mathrm{peak}}$, as
\[
T(n)=T_{\mathrm{single~atom}}+\gamma\,n_{\mathrm{peak}}.
\] 
We investigated this behavior by measuring the gas temperature for various atom densities. The atom density was varied by loading different atom numbers $4\times 10^7\leq N\leq 2\times10^8$  or using different gradient values $1.1\,\mathrm{G/cm}\leq G\leq 5.3\,$G/cm. Note that the highest atom density used for this study ($n_{\mathrm{peak}}\simeq1.1\times10^{11}\,$cm$^{-3}$) exceeds the maximum density $n_{\mathrm{max}}$ discussed in the main text as we use here larger magnetic field gradients. As shown in figure\,\ref{Fig_Temperature_Appendix}b, our measurements are compatible with a linear variation of the temperature with density, with a slope $\gamma=8(1)\times 10^{-11}\,\mu\mathrm{K}\,\mathrm{cm}^3$.  This value is comparable with the one obtained with Cs gray molasses \cite{boiron1996laser}. 

\section{MOT temperature in the spin-polarized regime\label{Appendix_Temperature_3D}}
In this section we give a more detailed description of the MOT temperature calculation and extend the analysis to the atom dynamics in  three spatial directions. We restrict the discussion to the spin-polarized regime.

In order to extract the damping coefficient, the radiative force can be expanded at the equilibrium position $\mathbf{r}_c$ as
\begin{equation*}
\mathbf{F}_{\mathrm{rad}}(\mathbf{r}_c,\mathbf{v})=-m\alpha \left(\frac{2}{3}v_x\ex+v_y\ey+v_z\ez\right)+\mathcal{O}(v^2),
\end{equation*}
where $\alpha$ was introduced in the main text, see equation\;(\ref{eq_alpha}). The anisotropy of the damping comes the specific geometry of our setup  (see figure \ref{Fig_Scheme}b).

The momentum diffusion tensor $D$ is calculated as the sum of the diffusion tensors $D^{\mathrm{abs}}$ and $D^{\mathrm{em}}$, associated with the stochastic absorption and spontaneous emission events, respectively. Taking into account the geometry of our experiment (see figure \ref{Fig_Scheme}b), we obtain the diffusion coefficients
\begin{eqnarray*}
D^{\mathrm{abs}}&=&\frac{1}{16\eta}\hbar^2k^2\Gamma
\left(
\begin{array}{ccc}
2&0&0\\
0&3&0\\
0&0&3\\
\end{array}
\right),\\
D^{\mathrm{em}}&=&\frac{1}{20\eta}\hbar^2k^2\Gamma
\left(
\begin{array}{ccc}
3&0&0\\
0&3&0\\
0&0&4\\
\end{array}
\right).
\end{eqnarray*}

The temperature is then obtained according to $k_BT=D/(m\alpha)$ and reveals weak anisotropy:
\begin{eqnarray*}
\left(
\begin{array}{c}
T_x\\
T_y\\
T_z\\
\end{array}
\right)
=
\frac{\eta s}{\sqrt{s(\eta-2)-1}}
\left(
\begin{array}{c}
33/120\\
27/120\\
31/120\\
\end{array}
\right)\frac{\hbar\Gamma}{k_B}.
\end{eqnarray*}

\section{Temperature increase due to the laser intensity noise\label{Appendix_noise}}
In this section we give more details on the calculation of the temperature increase due to  fluctuations of the cooling laser intensity, leading to a time-dependent saturation parameter $s(t)$. The main heating effect comes from  fluctuations of the trap center $z_c(t)$, which can be related to $s(t)$ using the equilibrium condition (\ref{eq_equilibrium}). 

\begin{figure}
\includegraphics{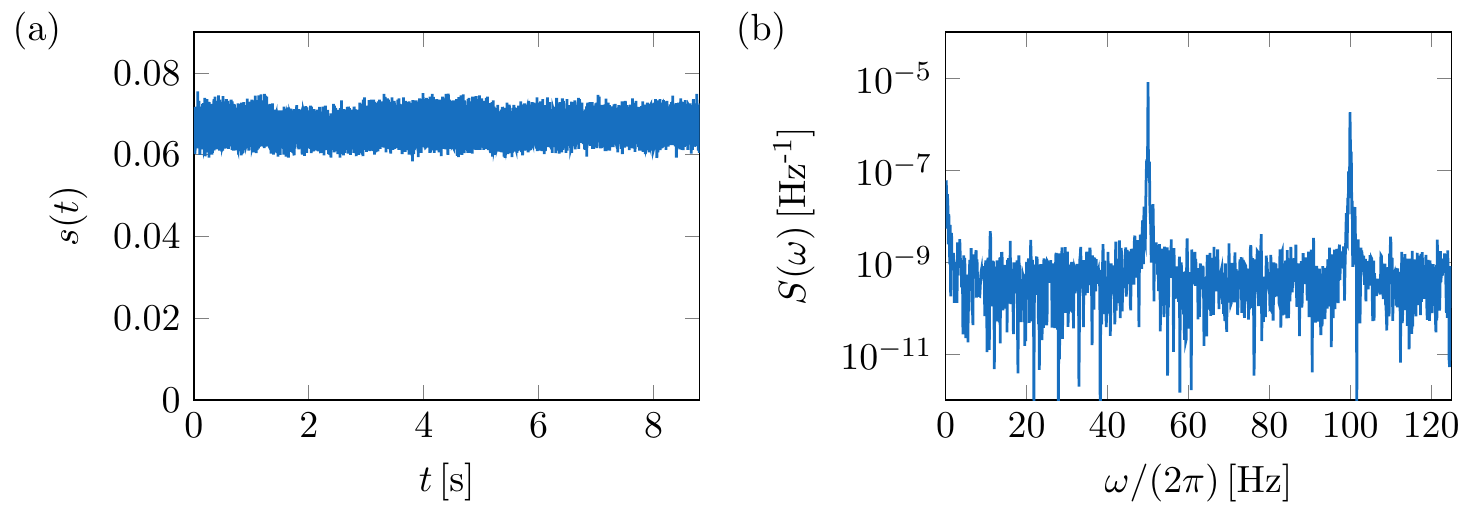}
\caption{
(a) Time evolution of the saturation parameter $s$, for the smallest mean value $s=0.065$ explored in figure \ref{Fig_Temperature}b. The r.m.s. fluctuations of $5\times10^{-3}$ result from the $\sim80$\,dB dynamic range of the intensity locking system. 
(b) Spectral density $S(\omega)$ of the saturation parameter noise corresponding to $(a)$.
\label{Fig_Noise}}
\end{figure}

For simplicity we restrict the discussion to the atom dynamics along the $z$ axis.
The atom dynamics is described by Newton's equation
\begin{equation}\label{eq_dyn}
m \ddot z=-\kappa_z[z-z_c(t)]-m\alpha\dot z+F_{\mathrm{L}}(t).
\end{equation}
Here, $F_{\mathrm{L}}(t)$ is the Langevin  force associated with stochastic radiative processes, such that $\left<F_{\mathrm{L}}(t)\right>=0$ and $\left<F_{\mathrm{L}}(t)F_{\mathrm{L}}(t')\right>=2D\delta(t-t')$, involving the diffusion coefficient introduced in the main text. By integrating equation (\ref{eq_dyn}), we calculate the r.m.s. fluctuations of the velocity, as
\begin{equation}\label{eq_sigmav}
\left<\dot z^2\right>=\frac{D}{m^2\alpha}+\left(\frac{\dd z_c}{\dd s}\right)^2\int\dd\omega\,\frac{\omega_0^4\omega^2}{(\omega_0^2-\omega^2)^2+\omega^2\alpha^2}S(\omega),
\end{equation}
where $S(\omega)$ is the spectral density of the saturation parameter noise and $\omega_0=\sqrt{\kappa_z/m}$.
The equilibrium temperature  is then obtained as $T=m\left<\dot z^2\right>/k_B$. Note that for small damping rates $\alpha$ and in the absence of Langevin forces, equation (\ref{eq_sigmav}) is consistent with the heating rates expected from reference \cite{savard1997laser} for shaken conservative traps. 
The dashed line in figure \ref{Fig_Temperature}b is calculated using equation (\ref{eq_sigmav}) and the measured noise spectrum, shown in figure \ref{Fig_Noise} for the lowest saturation parameter $s=0.065$ explored in figure \ref{Fig_Temperature}b. Note that, during the MOT loading and compression, the laser intensity is servo-locked to a PID controller, typically over the range $0.065\lesssim s\leq 50$. 

\section{Calculation of the inelastic  loss rate\label{Appendix_beta}}
In this section we describe the two-body loss model used in the main text to interpret the measured loss coefficients, adapting thereby standard treatments to the case of Dysprosium atoms \cite{gallagher1989exoergic,julienne1991cold}.

We assume the non-linear atom losses to be triggered by the light-assisted formation of molecules. Once the pair of atoms is excited to an attractive molecular state,  strong Van der Waals forces induce fast atom dynamics, leading to atom losses once the molecule is de-excited after spontaneous emission.

A precise modelling is a challenging task, as it requires calculating the 646 molecular potentials and the corresponding excitation amplitudes, taking into account the local magnetic field value and the orientation of atom pairs. We consider here a single effective molecular potential $V_{\mathrm{mol}}(r)=-\lambda\hbar\Gamma/(kr)^3$, of  $1/e$ lifetime $(\mu\Gamma)^{-1}$. For simplicity we consider  a uniform atom density $n=N/V$. The atom loss can be described by the equation
\begin{equation}\label{eq1}
\dot N=-\frac{n^2}{2}\int\dd\mathbf{r_1}\,\dd\mathbf{r_2}\,2\Gamma_{\mathrm{asso}}(|\mathbf{r_1}-\mathbf{r_2}|)P_{\mathrm{loss}}(|\mathbf{r_1}-\mathbf{r_2}|),
\end{equation}
where $\Gamma_{\mathrm{asso}}(r)$ is the rate of molecule formation for a pair of atoms of relative distance $r$, and $P_{\mathrm{loss}}(r)$ is the probability to lose this pair of atoms after de-excitation. The factor $\frac{1}{2}$ avoids double counting and the factor 2 accounts for the fact that each loss event corresponds to the loss of two atoms. Equation (\ref{eq1}) can be recast as
\begin{equation*}
\dot N=-\beta n N,\quad \beta=\int\dd\mathbf{r}\,\Gamma_{\mathrm{asso}}(r)P_{\mathrm{loss}}(r).
\end{equation*}

 Consider a pair of atoms in the MOT separated by the distance $r_{\mathrm{i}}$. We assume the rate of molecular association, triggered by the absorption of a photon ($q=-1$, 0 or 1 referring to $\sigma_-$, $\pi$ or $\sigma_+$ polarizations, respectively) by an atom of spin $\ket{J,m_J}$, to be given by the standard algebra of atom-light interaction, as
\begin{equation}\label{eq_Gamma_mol2}\nonumber
\Gamma_{\mathrm{asso}}(r_{\mathrm{i}})=
\left|
\left<J',m_J+q\middle | J,m_J;1,q\right>
\right|^2\frac{\mu\Gamma}{2}\frac{2s}{1+4[\Delta_{\mathrm{loc}}-V_{\mathrm{mol}}(r_{\mathrm{i}})/\hbar]^2/(\mu\Gamma)^2},
\end{equation}
where we ignore intensity saturation effects. 

Once the molecule is formed, the atom pair evolves according to Newton's law $(m/2)\ddot r=-\partial_r V_{\mathrm{mol}}(r)$, which can be solved implicitly. We neglect the initial atom motion as it corresponds to a weak energy scale for this problem. The electronic excitation decays after a duration $\tau$, corresponding to a distance $r_{\mathrm{f}}(r_{\mathrm{i}},\tau)$, such that
\[
\int_{r_{\mathrm{f}}}^{r_{\mathrm{i}}}\frac{\dd r}{\sqrt{r^{-3}-r_{\mathrm{i}}^{-3}}}=2\sqrt{\frac{\lambda\hbar\Gamma}{m k^3}}\tau.
\]
The acquired kinetic energy per atom $E_{\mathrm{k}}$ is obtained from energy conservation, as $2E_{\mathrm{k}}(r_{\mathrm{i}},\tau)=V(r_{\mathrm{i}})-V(r_{\mathrm{f}})$. 

The atom pair is lost if the acquired kinetic energy exceeds a threshold energy $E^*$. This energy can be calculated by solving numerically the equation of motion of an atom of kinetic energy $E^*$, initially located at the equilibrium position $\mathbf{r}_c$, and subjected to the radiative forces.  For the MOT parameters corresponding to the data of atom number represented in figure\;\ref{Fig_Atom_Losses}a, 
 we estimate a capture velocity $v^*$ of about 0.6\,m/s, corresponding to an energy $E^*\simeq 500\,\hbar\Gamma$.

The loss probability is then obtained as the probability for the atoms to acquire enough kinetic energy during the molecular dynamics: 
\[
P_{\mathrm{loss}}(r_{\mathrm{i}})
=\int_0^{\infty}\dd \tau\,\mu\Gamma\, e^{-\mu\Gamma\tau}\Theta\left[E_{\mathrm{k}}(r_{\mathrm{i}},\tau)-E^*\right],
\]
where $\mu\Gamma e^{-\mu\Gamma\tau}$ is the density probability for the spontaneous emission to occur at time $\tau$ ($\Theta$ is the Heaviside function). We obtain
\[
P_{\mathrm{loss}}(r_{\mathrm{i}})
=\exp\left(-\frac{\mu}{2\sqrt{\lambda}}(kr_{\mathrm{i}})^{5/2}\sqrt{\frac{\hbar\Gamma}{2\Er}}f\left[\left(1+\frac{ E^*}{|V_{\mathrm{mol}}(r_i)|}\right)^{-1/3}\right]\right),
\]
where $f(x)=\int_x^1\frac{\dd u}{\sqrt{u^{-3}-1}}$. As the threshold energy $E^*$ is much larger than other energy scales, we can safely replace the  factor $f[\cdot]$ by $f(0)=\sqrt{\pi}\Gamma_{\mathrm{E}}(5/6)/[3\,\Gamma_{\mathrm{E}}(4/3)]$.

The loss rate can then be obtained by calculating numerically the integral (\ref{eq1}). We find that replacing the Lorentz absorption profile (\ref{eq_Gamma_mol2}) by a strict resonance condition (i.e. the suitable Dirac $\delta$ function) only introduces minor differences for the numerical value of $\beta$. After this replacement the integral can be calculated analytically, leading to the formula (\ref{eq_beta}) in the main text.

\section*{References}
%

\begin{thebibliography}{10}

\bibitem{lu2011strongly}
M~Lu, N~Q Burdick, S~H Youn, and B~L Lev.
\newblock {Strongly dipolar Bose-Einstein condensate of dysprosium}.
\newblock {\em Phys. Rev. Lett.}, 107(19):190401, 2011.

\bibitem{aikawa2012bose}
K~Aikawa, A~Frisch, M~Mark, S~Baier, A~Rietzler, R~Grimm, and F~Ferlaino.
\newblock {Bose-Einstein condensation of erbium}.
\newblock {\em Phys. Rev. Lett.}, 108(21):210401, 2012.

\bibitem{lu2012quantum}
M~Lu, N~Q Burdick, and B~L Lev.
\newblock {Quantum degenerate dipolar Fermi gas}.
\newblock {\em Phys. Rev. Lett.}, 108(21):215301, 2012.

\bibitem{aikawa2014observation}
K~Aikawa, S~Baier, A~Frisch, M~Mark, C~Ravensbergen, and F~Ferlaino.
\newblock {Observation of Fermi surface deformation in a dipolar quantum gas}.
\newblock {\em Science}, 345(6203):1484, 2014.

\bibitem{baier2016extended}
S~Baier, M~J Mark, D~Petter, K~Aikawa, L~Chomaz, Z~Cai, M~Baranov, P~Zoller,
  and F~Ferlaino.
\newblock {Extended Bose-Hubbard models with ultracold magnetic atoms}.
\newblock {\em Science}, 352(6282):201, 2016.

\bibitem{ferrier2016observation}
I~Ferrier-Barbut, H~Kadau, M~Schmitt, M~Wenzel, and T~Pfau.
\newblock {Observation of quantum droplets in a strongly dipolar Bose gas}.
\newblock {\em Phys. Rev. Lett.}, 116(21):215301, 2016.

\bibitem{kadau2016observing}
H~Kadau, M~Schmitt, M~Wenzel, C~Wink, T~Maier, I~Ferrier-Barbut, and T~Pfau.
\newblock {Observing the Rosensweig instability of a quantum ferrofluid}.
\newblock {\em Nature}, 530(7589):194, 2016.

\bibitem{frisch2014quantum}
A~Frisch, M~Mark, K~Aikawa, F~Ferlaino, J~L Bohn, C~Makrides, A~Petrov, and
  S~Kotochigova.
\newblock Quantum chaos in ultracold collisions of gas-phase erbium atoms.
\newblock {\em Nature}, 507(7493):475, 2014.

\bibitem{maier2015emergence}
T.~Maier, H.~Kadau, M.~Schmitt, M.~Wenzel, I.~Ferrier-Barbut, T.~Pfau,
  A.~Frisch, S.~Baier, K.~Aikawa, L.~Chomaz, M.~J. Mark, F.~Ferlaino,
  C.~Makrides, E.~Tiesinga, A.~Petrov, and S.~Kotochigova.
\newblock {Emergence of chaotic scattering in ultracold Er and Dy}.
\newblock {\em Phys. Rev. X}, 5(4):041029, 2015.

\bibitem{cui2013synthetic}
X~Cui, B~Lian, T-L Ho, B~L Lev, and H~Zhai.
\newblock Synthetic gauge field with highly magnetic lanthanide atoms.
\newblock {\em Phys. Rev. A}, 88(1):011601, 2013.

\bibitem{nascimbene2013realizing}
S~Nascimbene.
\newblock Realizing one-dimensional topological superfluids with ultracold
  atomic gases.
\newblock {\em J. Phys. B: At., Mol. Opt. Phys.}, 46(13):134005, 2013.

\bibitem{burdick2016long}
N~Q Burdick, Y~Tang, and B~L Lev.
\newblock {\em Phys. Rev. X}, 6(3):031022, 2016.

\bibitem{youn2010dysprosium}
S~H Youn, M~Lu, U~Ray, and B~L Lev.
\newblock Dysprosium magneto-optical traps.
\newblock {\em Phys. Rev. A}, 82(4):043425, 2010.

\bibitem{lu2010trapping}
M~Lu, S~H Youn, and B~L Lev.
\newblock Trapping ultracold dysprosium: a highly magnetic gas for dipolar
  physics.
\newblock {\em Phys. Rev. Lett.}, 104(6):063001, 2010.

\bibitem{maier2014narrow}
T~Maier, H~Kadau, M~Schmitt, A~Griesmaier, and T~Pfau.
\newblock Narrow-line magneto-optical trap for dysprosium atoms.
\newblock {\em Opt. Lett.}, 39(11):3138, 2014.

\bibitem{hemmerling2014buffer}
B~Hemmerling, G~K Drayna, E~Chae, A~Ravi, and J~M Doyle.
\newblock {Buffer gas loaded magneto-optical traps for Yb, Tm, Er and Ho}.
\newblock {\em New J. Phys.}, 16(6):063070, 2014.

\bibitem{miao2014magneto}
J~Miao, J~Hostetter, G~Stratis, and M~Saffman.
\newblock Magneto-optical trapping of holmium atoms.
\newblock {\em Phys. Rev. A}, 89(4):041401, 2014.

\bibitem{mcclelland2006laser}
J~J McClelland and J~L Hanssen.
\newblock Laser cooling without repumping: a magneto-optical trap for erbium
  atoms.
\newblock {\em Phys. Rev. Lett.}, 96(14):143005, 2006.

\bibitem{berglund2008narrow}
A~J Berglund, J~L Hanssen, and J~J McClelland.
\newblock Narrow-line magneto-optical cooling and trapping of strongly magnetic
  atoms.
\newblock {\em Phys. Rev. Lett.}, 100(11):113002, 2008.

\bibitem{frisch2012narrow}
A~Frisch, K~Aikawa, M~Mark, A~Rietzler, J~Schindler, E~Zupani{\v{c}}, R~Grimm,
  and F~Ferlaino.
\newblock Narrow-line magneto-optical trap for erbium.
\newblock {\em Phys. Rev. A}, 85(5):051401, 2012.

\bibitem{sukachev2010magneto}
D~Sukachev, A~Sokolov, K~Chebakov, A~Akimov, S~Kanorsky, N~Kolachevsky, and
  V~Sorokin.
\newblock Magneto-optical trap for thulium atoms.
\newblock {\em Phys. Rev. A}, 82(1):011405, 2010.

\bibitem{kuwamoto1999magneto}
T~Kuwamoto, K~Honda, Y~Takahashi, and T~Yabuzaki.
\newblock {Magneto-optical trapping of Yb atoms using an intercombination
  transition}.
\newblock {\em Phys. Rev. A}, 60(2):R745, 1999.

\bibitem{katori1999magneto}
H~Katori, T~Ido, Y~Isoya, and M~Kuwata-Gonokami.
\newblock Magneto-optical trapping and cooling of strontium atoms down to the
  photon recoil temperature.
\newblock {\em Phys. Rev. Lett.}, 82(6):1116, 1999.

\bibitem{binnewies2001doppler}
T~Binnewies, G~Wilpers, U~Sterr, F~Riehle, J~Helmcke, T~E Mehlst{\"a}ubler, E~M
  Rasel, and W~Ertmer.
\newblock Doppler cooling and trapping on forbidden transitions.
\newblock {\em Phys. Rev. Lett.}, 87(12):123002, 2001.

\bibitem{gustavsson1979lifetime}
M~Gustavsson, H~Lundberg, L~Nilsson, and S~Svanberg.
\newblock Lifetime measurements for excited states of rare-earth atoms using
  pulse modulation of a cw dye-laser beam.
\newblock {\em J. Opt. Soc. Am.}, 69(7):984, 1979.

\bibitem{loftus2004narrow}
T~H Loftus, T~Ido, M~M Boyd, A~D Ludlow, and J~Ye.
\newblock Narrow line cooling and momentum-space crystals.
\newblock {\em Phys. Rev. A}, 70(6):063413, 2004.

\bibitem{lu2011spectroscopy}
M~Lu, S~H Youn, and B~L Lev.
\newblock Spectroscopy of a narrow-line laser-cooling transition in atomic
  dysprosium.
\newblock {\em Phys. Rev. A}, 83(1):012510, 2011.

\bibitem{leefer2010transverse}
N~Leefer, A~Cing{\"o}z, B~Gerber-Siff, A~Sharma, J~R Torgerson, and D~Budker.
\newblock Transverse laser cooling of a thermal atomic beam of dysprosium.
\newblock {\em Phys. Rev. A}, 81(4):043427, 2010.

\bibitem{metcalf2007laser}
Harold~J Metcalf and Peter van~der Straten.
\newblock {\em Laser cooling and trapping of neutral atoms}.
\newblock Springer New York, 1999.

\bibitem{wineland1979laser}
D~J Wineland and W~M Itano.
\newblock Laser cooling of atoms.
\newblock {\em Phys. Rev. A}, 20(4):1521, 1979.

\bibitem{drewsen1994investigation}
M~Drewsen, P~Laurent, A~Nadir, G~Santarelli, A~Clairon, Y~Castin, D~Grison, and
  C~Salomon.
\newblock {Investigation of sub-Doppler cooling effects in a cesium
  magneto-optical trap}.
\newblock {\em Appl. Phys. B}, 59(3):283, 1994.

\bibitem{boiron1996laser}
D~Boiron, A~Michaud, P~Lemonde, Y~Castin, C~Salomon, S~Weyers, K~Szymaniec,
  L~Cognet, and A~Clairon.
\newblock {Laser cooling of cesium atoms in gray optical molasses down to 1.1
  $\mu$K}.
\newblock {\em Phys. Rev. A}, 53(6):R3734, 1996.

\bibitem{kerman2000beyond}
A~J Kerman, V~Vuleti{\'c}, C~Chin, and S~Chu.
\newblock {Beyond optical molasses: 3D Raman sideband cooling of atomic cesium
  to high phase-space density}.
\newblock {\em Phys. Rev. Lett.}, 84(3):439, 2000.

\bibitem{walker1990collective}
T~Walker, D~Sesko, and C~Wieman.
\newblock Collective behavior of optically trapped neutral atoms.
\newblock {\em Phys. Rev. Lett.}, 64(4):408, 1990.

\bibitem{townsend1995phase}
C~G Townsend, N~H Edwards, C~J Cooper, K~P Zetie, C~J Foot, A~M Steane,
  P~Szriftgiser, H~Perrin, and J~Dalibard.
\newblock Phase-space density in the magneto-optical trap.
\newblock {\em Phys. Rev. A}, 52(2):1423, 1995.

\bibitem{gallagher1989exoergic}
A~Gallagher and D~E Pritchard.
\newblock {Exoergic collisions of cold Na$^*$-Na}.
\newblock {\em Phys. Rev. Lett.}, 63(9):957, 1989.

\bibitem{julienne1991cold}
P~S Julienne and J~Vigu{\'e}.
\newblock Cold collisions of ground- and excited-state alkali-metal atoms.
\newblock {\em Phys. Rev. A}, 44(7):4464, 1991.

\bibitem{julienne1992theory}
P~S Julienne, A~M Smith, and K~Burnett.
\newblock Theory of collisions between laser cooled atoms.
\newblock {\em Adv. At. Mol. Opt. Phys.}, 30:141, 1992.

\bibitem{dinneen1999cold}
T~P Dinneen, K~R Vogel, E~Arimondo, J~L Hall, and A~Gallagher.
\newblock {Cold collisions of Sr$^*$- Sr in a magneto-optical trap}.
\newblock {\em Phys. Rev. A}, 59(2):1216, 1999.

\bibitem{lett1995hyperfine}
P~D Lett, K~M{\o}lmer, S~D Gensemer, K~Y~N Tan, A~Kumarakrishnan, C~D Wallace,
  and P~L Gould.
\newblock Hyperfine structure modifications of collisional losses from
  light-force atom traps.
\newblock {\em J. Phys. B: At., Mol. Opt. Phys.}, 28(1):65, 1995.

\bibitem{hensler2003dipolar}
S~Hensler, J~Werner, A~Griesmaier, P~O Schmidt, A~G{\"o}rlitz, T~Pfau,
  S~Giovanazzi, and K~Rz{\c a}{\.z}ewski.
\newblock Dipolar relaxation in an ultra-cold gas of magnetically trapped
  chromium atoms.
\newblock {\em Appl. Phys. B}, 77(8):765, 2003.

\bibitem{burdick2015fermionic}
N~Q Burdick, K~Baumann, Y~Tang, M~Lu, and B~L Lev.
\newblock Fermionic suppression of dipolar relaxation.
\newblock {\em Phys. Rev. Lett.}, 114(2):023201, 2015.

\bibitem{berglund2007sub}
A~J Berglund, S~A Lee, and J~J McClelland.
\newblock {Sub-Doppler laser cooling and magnetic trapping of erbium}.
\newblock {\em Phys. Rev. A}, 76(5):053418, 2007.

\bibitem{youn2010anisotropic}
S~H Youn, M~Lu, and B~L Lev.
\newblock {Anisotropic sub-Doppler laser cooling in dysprosium magneto-optical
  traps}.
\newblock {\em Phys. Rev. A}, 82(4):043403, 2010.

\bibitem{dalibard1989laser}
J~Dalibard and C~Cohen-Tannoudji.
\newblock {Laser cooling below the Doppler limit by polarization gradients:
  simple theoretical models}.
\newblock {\em J. Opt. Soc. Am. B}, 6(11):2023, 1989.

\bibitem{ungar1989optical}
P~J Ungar, D~S Weiss, E~Riis, and S~Chu.
\newblock Optical molasses and multilevel atoms: theory.
\newblock {\em J. Opt. Soc. Am. B}, 6(11):2058, 1989.

\bibitem{savard1997laser}
T~A~Savard, K~M~O'Hara, and J~E~Thomas.
\newblock Laser-noise-induced heating in far-off resonance optical traps.
\newblock {\em Phys. Rev. A}, 56(2):R1095, 1997.

\end{thebibliography}

\end{document}